\documentclass[twocolumn,aps,floatfix]{revtex4}
\usepackage{amssymb}
\usepackage{graphicx}

\begin{document}
%\draft

\title{Splitting of doubly quantized vortices in dilute Bose-Einstein 
       condensates}

\author{K. Gawryluk,$\,^1$ T. Karpiuk,$\,^1$ M. Brewczyk,$\,^1$ 
        and K. Rz{\c a}\.zewski$\,^{2,3}$}

\affiliation{\mbox{$^1$ Wydzia{\l} Fizyki, Uniwersytet w Bia{\l}ymstoku,
                        ulica Lipowa 41, 15-424 Bia{\l}ystok, Poland}  \\
\mbox{$^2$ Centrum Fizyki Teoretycznej PAN, Aleja Lotnik\'ow 32/46, 02-668 Warsaw,
           Poland}   
\mbox{$^3$ Faculty of Mathematics and Sciences UKSW, Warsaw, Poland}    }

%\date{\today}

\begin{abstract}

We investigate the dynamics of doubly charged vortices generated in dilute Bose-Einstein 
condensates by using the topological phase imprinting technique. We find splitting
times of such vortices and show that thermal atoms are responsible for their decay.

\end{abstract}

\maketitle

In their recent Letter \cite{m2Virtanen}, Huhtam\"aki {\it et al.} theoretically
investigated the splitting of a topologically imprinted doubly charged vortex into
two singly charged vortices as occurring in a dilute atomic Bose-Einstein condensate.
They compare the results of simulation with recent experiment \cite{Ketterle} and
show that the combination of gravitational sag and the time dependence of the 
trapping potential alone are enough to explain the observed splitting times.
Based on such an outcome the authors of Ref. \cite{m2Virtanen} claim that, contrary to
previous theoretical results \cite{m2}, the thermal excitations are not relevant in
modeling the experiment of Ref. \cite{Ketterle}. 
We are going to show in this Brief Report that, indeed, the opposite is true.
In fact, a number of thermal (uncondensed) atoms appears in the system while disturbing the 
gas. They continue to appear after the perturbation is over and until the doubly quantized
vortex breaks into two singly quantized vortices.
The overall number of uncondensed atoms remains approximately on the level of $20\%$, 
which is already at the edge of experimental detection capabilities. However, the uncondensed
atoms do not form the broad cloud allowing the identification by fitting to a bimodal
distribution - they are rather located in the core of the vortex and therefore are harder 
to detect.
Perhaps, the signatures of the presence of thermal atoms in vortices cores are already
visible in experiment in a way that after splitting the cores of two singly
charged vortices get darker in comparison with the core of initially imprinted doubly
quantized vortex (see Fig. 2 in Ref. \cite{Ketterle}).

To investigate the thermal excitations in a Bose gas we use the classical 
fields approximation \cite{przeglad} - an approach that treats both condensed and thermal 
atoms at the same footing until the detection time when the splitting into 
the condensate and the thermal cloud occurs.
Technically, such a decomposition requires calculation of time and space average
of a one-particle density matrix built of the classical field evolving according to
the Gross-Pitaevskii equation \cite{przeglad}.

Therefore, we have repeated the calculations of Ref. \cite{m2Virtanen}. As in
\cite{m2Virtanen}, we closely follow the experiment reported in \cite{Ketterle}
and solve the Gross-Pitaevskii equation with time-dependent trapping potential (according 
to the idea of topological phase imprinting \cite{Isoshima}) combined with the gravitational 
potential. However, we interpret the solution of the Gross-Pitaevskii equation as a classical 
field and by using the space averaging procedure we determine the condensate and the thermal 
cloud at each time.

\begin{figure}[tbh]
\resizebox{3.0in}{2.1in}
{\includegraphics{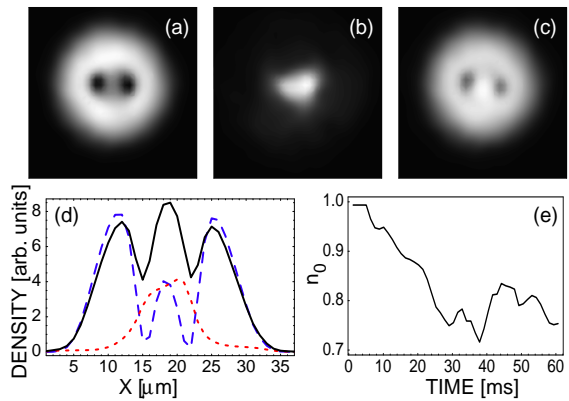}}
\caption{(Color online). The $z$-integrated (over [$-15$ $\mu$m, $15$ $\mu$m] interval)
condensate (a), thermal (b), and total (c) densities at $47$\,ms for the interaction 
strength $an_z=8.5$; the horizontal cuts of the condensate (dashed line), thermal (dotted 
line), and total (solid line) densities (d); the condensate fraction as a function of 
time (e).}
\label{thermal}
\end{figure}

Fig. \ref{thermal} clearly shows that the uncondensed atoms appear in the system during 
the evolution. Although initially all atoms are in the condensate (i.e., the condensate
fraction equals $1$ as in Fig. \ref{thermal}(e)), already after $6$\,ms distinguishable
fraction of uncondensed atoms is produced. This is not surprising since the process of
imprinting the vortex is accompanied by a sudden squeeze of the condensate in the radial 
direction and a kick of it in the vertical direction \cite{Ketterle}.
The thermal atoms continue to appear after the imprinting is over and until the doubly 
quantized vortex splits into two singly charged vortices
(production of thermal atoms while the system was initially at zero temperature was
also reported in Ref. \cite{Lobo}, where the crystallization of a vortex lattice in a
stirred Bose-Einstein condensate was investigated).

Fig. \ref{thermal} also proves that the thermal atoms are mainly located in the vortex core. 
On the other hand, the thermal noise on the level of about $20\%$ leads to the decay of a 
vortex in approximately $45$\,ms (see Fig. 3 in Ref. \cite{m2}) in agreement with the 
calculations in \cite{m2Virtanen} and this Brief Report. We claim this means that, indeed, 
behind the origin of the splitting of the doubly quantized vortex reported in \cite{m2Virtanen} 
(the impetus given by the time dependence of the external potential) lies the presence of 
uncondensed atoms appearing in the system as a result of its disturbance. 
In other words, the understanding of the experiment by Shin {\it et al.} requires going beyond 
the standard Gross-Pitaevskii approximation. It is clear that other kinds of perturbation 
of the condensate (for instance as the one considered in Ref. \cite{Spain}) will result in
a production of uncondensed atoms and consequently lead to the decay of the multiply charged
vortex.

In conclusion, we have studied the splitting process of doubly charged vortices created 
via the topological phase imprinting method. We found that the uncondensed atoms,
inevitably produced in the condensate while it is disturbed, are responsible for the
decay of such vortices. This is because the unstable modes, located in the vortex core, 
are seeded by the uncondensed atoms which are also located there.

We thank M. Gajda for helpful discussions. 
We acknowledge support by the Polish KBN Grant No. 1 P03B 051 30
and by Polish Government research funds for 2006-2009.  \\

\end{document}